# A Novel Fuzzy Logic-Based Metric for Audio Quality Assessment

Objective Audio Quality Assessment

Luis F. Abanto-Leon

Department of Access Technologies
and Radiopropagation
INICTEL – UNI / Peruvian
University of Applied Sciences
Lima, Peru
u510950@upc.edu.pe

Guillermo Kemper Vásquez

Department of Access Technologies
and Radiopropagation
INICTEL - UNI
Lima, Peru
glkemper@hotmail.com

Joel Telles Castillo

Department of Access Technologies
and Radiopropagation
INICTEL - UNI
Lima, Peru
jtelles@inictel-uni.edu.pe

*Abstract*— ITU-R BS.1387 states a method for objective assessment of perceived audio quality. This Recommendation, known also as PEAQ (Perceptual Evaluation of Audio Quality) is based on a psychoacoustic model of the human ear and was standardized by the International Telecommunications Union as an alternative to subjective tests, which are expensive and time-consuming processes. PEAQ combines various physiological and psycho-acoustical properties of the human ear to give a measure of the quality difference between a reference audio and a test audio. The reference audio signal could be considered as a distortion-free source, whereas the test signal is a distorted version of the reference, which may have audible artifacts because of compression. The algorithm computes the Model Output Variables (MOVs) which are mapped to a single quality measure, Objective Difference Grade (ODG), using a three-layer perceptron artificial neural network. The ODG estimates the perceived distortion between both audio signals. In this paper we propose a new metric of low computational complexity called FQI (Fuzzy Quality Index) which is based on Fuzzy Logic reasoning and has been incorporated into the existing PEAQ model to improve its overall performance. Results show that the modified version slightly outperforms PEAQ.

I. INTRODUCTION

The issue of digital audio compression has been of great interest to researchers and to the telecommunication industry since the digital revolution began in 1970s [2], when Pulse Code Modulation was first introduced for digital recording.

Nowadays, compression algorithms have become the state-of-the-art technology in modern telecommunication and embedded systems that transmit or store audio digitally [1]. Compression algorithms can be found in advanced telecommunication systems such as Digital Television as well as in less complex systems such as MP3 Audio Players.

The fundamental issue in digital audio compression is to fit or transmit a large amount of information into a reduced storage space or limited bandwidth, while maintaining the best possible audio quality. By compressing audio at high bitrates, annoying perceptible artifacts have low occurrence, resulting in high quality audio. On the other hand, when compression is performed at low bitrates the occurrence of audible artifacts is high, resulting in low quality audio. Faced with this problem, how could we find out if an audio signal has good or bad quality? The most common approach for determining the quality of audio signals is through subjective tests. But subjective tests are impractical because their implementation demand trained listeners to assess the quality of audio signals [3]. In replace of subjective tests, objective computer-based measurement methods are an appropriate solution, because they can estimate the perceived audio quality without the intervention of human listeners.

We have fully implemented ITU-R BS.1387 Recommendation in two programming environments: C# and Matlab. On the basis of these implementations we have developed a new quality metric based on fuzzy logic reasoning.

The remainder of this paper is organized as follows. In the next section we describe the methodology that we followed to perform the subjective tests. In section III, we explain the PEAQ algorithm and its stages of processing. In section IV we present a brief background about Fuzzy Logic and explain our metric. In section V, we describe the ANN characteristics and the training algorithm we used. In section VI, results of experiments are showed. Finally in section VII, conclusions of this investigation are summarized.

II. AUDIO QUALITY EVALUATION

*A. Subjective Tests*

Subjective assessment was performed according to ITU-R Recommendation BS.1116 [4] and was executed with the aim of obtaining the Subjective Difference Grade (SDG) values for a large audio database. In the tests, the listeners were presented to three signals A, B and C. Signal A was the unprocessed reference audio signal, B and C were either the signals under test or a copy of the reference A (hidden reference). Listeners were asked to score the impairments (audible artifacts) of B and C with respect to signal A on a continuous impairment scale that ranges from 1 to 5 (see Fig. 1).

The final quality value, SDG, is given by the difference of listeners' ratings for signals B and C. The SDG is defined as

$$SDG = Grade_{Test} - Grade_{Reference} \quad (1)$$

The Subjective Difference Grade ideally ranges from -4 to 0. A value of -4 represents a very annoying impairment whereas a value of 0 means an imperceptible impairment.

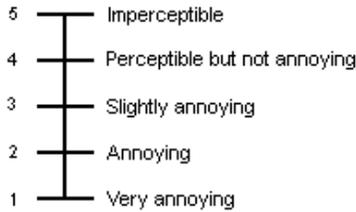

Fig. 1. ITU-R BS.1116 Impairment Scale

Subjective tests were carried out using audio signals with durations between 15 and 35 seconds. These tests were performed with the subjects staying alone in an acoustic anti-echoic room listening through headphones to the audio signals. EBU SQAM Audio Database [18] was selected to perform the subjective tests. In the first test that was performed, 8 original audio signals were selected to create 10 processed samples from each audio using one of the following four codecs LAME MP3, MPEG4-AAC, WMA and OGG comprising a wide range of bitrates. This first dataset was exclusively used for training the ANN. For the second test, 10 audio files were selected and processed under different bitrates using the aforementioned codecs. This last dataset was reserved for testing. Audio samples were assessed by 22 volunteers who scored the quality through a software application.

### B. Objective Tests

Objective tests differs from subjective tests in the fact that these methods are computer-based and do not depend on human listeners to assess the quality of audio. PEAQ is a Full Reference (FR) method, i.e. reference source is necessary to perform the evaluation of quality. On the contrary No-Reference (NR) methods do not need the reference source, but experiences had shown that FR methods correlate better with the subjective opinion.

The general block-diagram of PEAQ is illustrated in Fig. 2.

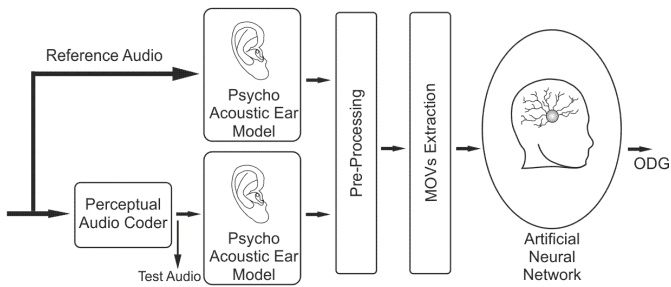

Fig. 2. PEAQ block-diagram

PEAQ method can be divided into four key stages: a psychoacoustic ear model, a pre-processing stage, a metrics extraction stage and a cognitive model in which a neural network maps the MOVs into a single quality measure.

### III. PEAQ Algorithm

In 1994 the International Telecommunications Union created a task group whose goal was to develop a Recommendation for objectively measuring perceived audio quality. An open call of proposals was issued and six perceptual measurement methods were presented : Disturbance Index (DIX) [5], Noise-to-Mask Ratio (NMR) [6], Perceptual Audio Quality Measure (PAQM) [7], Perceptual Evaluation (PERCEVAL) [8], Perceptual Objective Measure (POM) [9], and the Toolbox Approach [10] [11] [13] [15]. The best features of the original methods were combined into a new objective model for evaluating sound quality called Perceptual Evaluation of Audio Quality (PEAQ), and formally standardized as ITU-R Recommendation BS.1387 [11].

PEAQ specifies two different versions: the basic approach which employs the Short-Time Fourier Transform (STFT) to perform the time-frequency analysis and the advanced approach which uses a filter bank. In this paper we only discuss the basic version. The block-diagram of PEAQ's Psychoacoustic Ear Model is illustrated in Fig. 3.

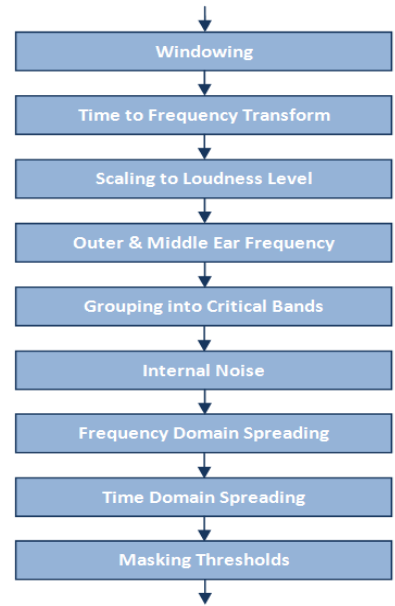

Fig. 3. Psychoacoustic Ear Model of PEAQ

Each of the parts that form PEAQ Ear Model will be explained in the following lines.

### A. Windowing

Audio signals are cut into frames of 42.6ms with an overlapping of 50%. Then, a Hanning window is applied to each of the resulting frames. A scaled Hanning window is shown in (2)

$$h[n] = \sqrt{\frac{8}{3}} \frac{1}{2}\left(1 - \cos\left(\frac{2\pi n}{N-1}\right)\right) \quad 0 \leq n \leq N-1 \quad (2)$$

where $N=2048$ represents the frame length; the overlapping is of 1024 samples. Equation (2) differs from a conventional Hanning window in the factor $\sqrt{8/3}$, which is justified by an energy compensation issue. A frame of the audio signal is represented by $x[n]$ and the windowed frame is represented by $x_w[n]$. Both signals are related by (3)

$$x_w[n] = x[n] \cdot h[n] \quad (3)$$

### B. Discrete Fourier Transform

The windowed signal is transformed to the frequency domain using the Discrete Fourier Transform (DFT).

$$F_f[k] = \frac{1}{N} \sum_{n=0}^{N-1} x_w[n] e^{-j\frac{2\pi}{N}kn} \quad (4)$$

In (4), $k$ and $n$ range from $0$ to $N-1$.

### C. Scaling Factor

The scaling factor for the FFT output is calculated from a full scale sine wave with Sound Pressure Level (SPL) $L_p$.

The normalization factor *Norm* is calculated by computing the maximum absolute value of the spectral coefficients over 10 frames, when using a sine wave of 1019.55 Hz and 0dB full scale (16 bits) as input signal.

$$fac = \frac{10^{L_p/20}}{Norm} \quad (5)$$

Next, the scaled FFT output is computed as follows

$$F[k] = fac \cdot F_f[k] \quad (6)$$

Subjective loudness (measured in sones) is related to the loudness level (measured in phones), which is in turn related to the sound pressure level [12]. So this stage is necessary in order to transform the input signals to a subjective loudness scale.

### D. Outer and Middle Ear Frequency Response

The outer ear is composed of the pinna and the auditory canal. The most important contribution of the pinna is that it increases the sound pressure level about 5dB for frequencies between 2 KHz to 3 KHz. The auditory canal has an intrinsic resonance frequency that increases the sound pressure about 3KHz at the eardrum [13].

The middle ear acts as an impedance-matching device between the outer ear and the inner ear, because the middle ear has the duty of transforming the sound waves in the air into waves propagated through the liquid that is contained in the cochlea [14].

The overall influence of these organs, outer and middle ear, can be modeled by the transfer function defined in (7)

$$W_d[k] = 10^{\frac{-2.184\left(\frac{f[k]}{1000}\right)^{-0.8} + 6.5 e^{-0.6\left(\frac{f[k]}{1000} - 3.3\right)^2} - 0.001\left(\frac{f[k]}{1000}\right)^{3.6}}{20}} \quad (7)$$

$$f[k] = k \cdot Fres \quad Fres = Fs/2048 \quad 0 \leq k \leq 1024$$

$W_d$ emulates the human ear sensitivity function to different frequencies. The weighted FFT output can be computed through (8)

$$F_e[k] = |F| \cdot W_d \quad (8)$$

### E. Critical bands

The organ of Corti has the function of transforming the mechanical oscillations within the inner ear into electrical signals that are transmitted to the auditory brainstem and to the auditory cortex. The duty of transforming mechanical waves into electrical signals is performed by groups of sensory cells which are represented by a filter bank that maps the DFT bins to 109 critical bands.

The pseudocode to perform this stage of processing can be found in [11]. The output of this process is $P_e$, which represents the *Pitch Mapped Energies*.

### F. Internal Noise

Noise that is present in the auditory nerve in addition to the noise caused by the flow of blood, are modeled by (9)

$$P_{Thres}[k] = 10^{0.1456\left(\frac{f_c[k]}{1000}\right)^{-0.8}} \quad (9)$$

where $f_c$ represents the central frequency of each of the 109 critical bands. Then, the internal noise is added to the *Pitch Mapped Energies* ($P_e$).

$$P_p[k,n] = P_e[k,n] + P_{Thres}[k] \quad (10)$$

The output of this stage of processing is $P_p$, *Pitch Patterns*.

### G. Frequency Spreading

The Pitch Patterns are smeared out over frequency using a spreading function. The spreading function is a two sided exponential. The lower slope is always 27dB/Bark and the upper slope is frequency and energy dependant.

$$S_u[k,n] = -24 - \frac{230}{f_c[k]} + 0.2L[k,n] \quad (11)$$

$$S_l[k,n] = 27 \quad (12)$$

$S_u$ represents the upper slope and $S_l$ represents the lower slope. $L$ is computed as the logarithm of the *Pitch Patterns* as shown in (13)

$$L[k,n] = 10 \cdot \log_{10}(P_p[k,n]) \qquad (13)$$

The output of this stage is $E_2$, called hereafter as *Unsmeared Excitation Patterns*, and expressed as shown in (14)

$$E_2[k,n] = \frac{1}{Norm_{SP}[k]} \left( \sum_{j=0}^{Z-1} E_{line}[j,k,n]^{0.4} \right)^{\frac{1}{0.4}} \qquad (14)$$

$E_{line}$ is expressed in (15) and assume that $res=0.25$ corresponds to the resolution of the pitch scale in Bark.

$$E_{line}[j,k,n] = \begin{cases} \dfrac{10^{\frac{L[j,n]}{10}} \cdot 10^{\frac{-res(j-k)S_l}{10}}}{\sum_{u=0}^{j-1} 10^{\frac{-res(j-u)S_l}{10}} + \sum_{u=j}^{Z-1} 10^{\frac{res(u-j)S_u[j,n]}{10}}} & \text{if } k < j \\[2em] \dfrac{10^{\frac{L[j,n]}{10}} \cdot 10^{\frac{res(k-j)S_u[j,n]}{10}}}{\sum_{u=0}^{j-1} 10^{\frac{-res(j-u)S_l}{10}} + \sum_{u=j}^{Z-1} 10^{\frac{res(u-j)S_u[j,n]}{10}}} & \text{if } k \geq j \end{cases} \qquad (15)$$

$Norm_{SP}$ is computed by (16)

$$Norm_{SP}[k] = \left( \sum_{j=0}^{Z-1} \tilde{E}_{line}[j,k]^{0.4} \right)^{\frac{1}{0.4}} \qquad (16)$$

where $\tilde{E}_{line}$ is computed by (17)

$$\tilde{E}_{line}[j,k,n] = \begin{cases} \dfrac{10^{\frac{-res(j-k)S_l}{10}}}{\sum_{u=0}^{j-1} 10^{\frac{-res(j-u)S_l}{10}} + \sum_{u=j}^{Z-1} 10^{\frac{res(u-j)S_u[j,0]}{10}}} & \text{if } k < j \\[2em] \dfrac{10^{\frac{res(k-j)S_u[j,0]}{10}}}{\sum_{u=0}^{j-1} 10^{\frac{-res(j-u)S_l}{10}} + \sum_{u=j}^{Z-1} 10^{\frac{res(u-j)S_u[j,0]}{10}}} & \text{if } k \geq j \end{cases} \qquad (17)$$

### H. Time Domain Spreading

This stage simulates the temporal masking effect, which is also known as non-simultaneous masking. There are two types of non-simultaneous masking, backward masking and forward masking, but only forward masking has been considered in this version. *Unsmeared Excitation Patterns* are smeared out in the time domain using a low pass filter as shown in (18)

$$E[k,n] = \max(a \cdot E[k,n-1] + (1-a) \cdot E_2[k,n], E_2[k,n]) \qquad (18)$$

Time constants are given by (19)

$$\tau = t_{min} + \frac{100}{fc[k]}(\tau_{100} - t_{min}) \qquad (19)$$

$$\tau_{100} = 0.030s \qquad \tau_{min} = 0.008$$

and $a$ is given by (20)

$$a = e^{-\frac{4}{187.5}\frac{1}{\tau}} \qquad (20)$$

The output of this stage is $E$ and is referred as *Excitation Patterns*.

### I. Masking Thresholds

*Masking Thresholds* are obtained by combining the *Excitation Patterns* with the weighting function $m$.

$$M[k,n] = \frac{E[k,n]}{10^{\frac{m[k]}{10}}} \qquad (21)$$

where $m$ is defined by (22)

$$m[k] = \begin{cases} 3.0 & \text{if } k \cdot res \leq 12 \\ 0.25 \cdot k \cdot res & \text{if } k \cdot res > 12 \end{cases} \qquad (22)$$

The output of this processing stage is $M$, referred as *Mask Patterns*.

### J. Pre-processing of excitation patterns

This stage performs level calibration, level correction and filtering of the reference and test processed patterns, improving the correlation between the test signal and the reference, in order to obtain more reliable comparisons. This stage can be divided in four parts.

- Level and pattern adaption: The average levels of Reference and Test Signals are adapted to each other using filters and level correction factors in order to compensate the distortions and level differences.

- Modulation: A measure for the modulation of the envelope at each output filter is calculated from the Simplified Loudness Patterns.

- Loudness: The specific loudness patterns are calculated and averaged across all filter channels.

- Error Signal: It is calculated by taking the absolute value of the difference between the Outer/Middle Ear Weighted FFT of the reference and test signal.

### K. Calculation of Model Output Variables

A number of 11 MOVs are calculated in base of the processed patterns of the previous stage. These metrics are shown in TABLE I and represent important properties of the human ear.

TABLE I. MODEL OUTPUT VARIABLES

| Model Output Variable | Description |
|---|---|
| WinModDiff1$_B$ | Windowed Modulation Difference |
| AvgModDiff1$_B$ | Average Modulation Difference |
| AvgModDiff2$_B$ | Average Modulation Difference |
| RmsNoiseLoud$_B$ | RMS value of the Distortion Loudness |
| BandwidthRef$_B$ | Bandwidth of the Reference Signal |
| BandwidthTest$_B$ | Bandwidth of the Test Signal |
| Total NMR$_B$ | Noise-to-Mask Ratio |
| RelDistFrames$_B$ | Fraction of frames with audible distortions |
| MFPD$_B$ | Maximum Filtered Probability of Detection |
| ADB$_B$ | Average Block Distortion |
| EHS$_B$ | Harmonic Structure of the Error |

For deeper comprehension of the aforementioned perceptual metrics, readers are encouraged to consult [11], where these metrics are explained in detail.

*L. ODG Calculation*

The ODG is obtained by providing the eleven metrics as inputs to the neural network as illustrated in Fig. 4.

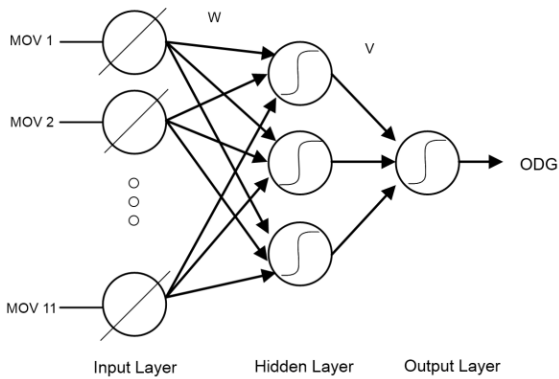

Fig. 4. Neural Network used in PEAQ

The ANN is composed if 11 linear neurons at the input, 3 sigmoidal neurons at the hidden layer and a single sigmoidal neuron at the output. Trained weights *W* and *V* can be found in [11].

## IV. A NEW METRIC

We propose a new fuzzy logic-based metric to be incorporated to the existing eleven that were mentioned in the previous section. We have employed fuzzy logic because it has the ability to mimic the human reasoning and has important advantage over hard-computing when dealing with non exact data. In this regards, we think that FL can deal with the vagueness that is involved in the assessment of audio quality.

*A. Level Adaption of Pitch Patterns*

As the first step, we proceed to compute a temporal factor for level correction as shown in (23).

$$K[n] = \left( \frac{\sum_{k=0}^{Z-1} \sqrt{Pp_{Test}[k,n] \cdot Pp_{Ref}[k,n]}}{\sum_{k=0}^{Z-1} Pp_{Test}[k,n]} \right)^2 \quad (23)$$

Where *n* represents the number of frame and *Z* represents the total number of critical bands, Z=109.

$$NPp_{Ref}[k,n] = K[n] \cdot Pp_{Ref}[k,n] \quad if \quad K[n] > 1 \quad (24)$$

$$NPp_{Test}[k,n] = K[n] \cdot Pp_{Test}[k,n] \quad if \quad K[n] \leq 1 \quad (25)$$

Level-compensated Pitch Patterns, $NPp_{Ref}$ and $NPp_{Test}$, are obtained as shown in (24) and (25) according to the value that *K[n]* takes. This step is necessary in order to adapt the Reference Pitch Pattern and the Test Pitch Pattern to each other so that a fairer comparison between the two patterns can be achieved. We have chosen Pitch Patterns to be the basis of the proposed metric because they are the result of mapping mechanical oscillation into electrical signals (energies) to a critical band scale, which is related to the hearing sensations. This approach is based in the fact that our hearing system analyses a broad spectrum into parts that correspond to critical bands. It is also worth saying that the critical band scale is closely related to other scales that describe characteristics of the hearing system, e.g. just-noticeable frequency variations are related to critical bandwidth [14].

*B. Pitch Patterns Smoothing*

A simplified envelope over the critical bands domain is computed through the use of a first order digital filter as shown in (26)

$$F[k,n] = (1-\beta) \cdot F[k,n-1] + \beta \cdot \frac{\sum_{i=-W_1}^{W_2} \alpha[W_1+i] \cdot P[k+i,n]}{\sum_{i=-W_1}^{W_2} \alpha[W_1+i]} \quad (26)$$

Consider that $W_1$ and $W_2$ are the lower and upper limits for the window.

$$W_1 = \max(0, k-U) \quad (27)$$

$$W_2 = \min(N-1, n+U) \quad (28)$$

In (29), *α[i]* represents the weighting coefficients of the triangular window.

$$\alpha[i] = \begin{cases} i+1 & if \quad i \leq U+1 \\ 2 \cdot U - i + 1 & if \quad i > U+1 \end{cases} \quad (29)$$

Consider that *N* is the total number of frames and *U*, which is related to the window size, has been set to 3. In our experiments, β was set to 0.85. In (26), assume that *P[k,n]* represents the input, i.e. the Level-compensated Pitch Patterns, which are obtained from (24) and (25). $F[k,n]$ represents the smoothed output referred as *Filtered Pitch Patterns* and obtained from (26). This stage is important because smoothing reduces the sensitivity to short-term high variations.

*C. Filtered Pitch Patterns Post-processing*

In base of the results that were obtained in the previous stage, we compute $D_p[k,n]$ which represents a simple level-based similarity index. Values that $D_p[k,n]$ takes depend on the evaluation of the two conditions expressed in (30). These conditions are aimed to assign $D_p[k,n]$ the value, between two, that best meets the requirement of minimum error.

$$D_p[k,n] = \begin{cases} \dfrac{FP_R[k,n]}{FP_T[k,n]} & \text{if } \left|1-\dfrac{FP_R[k,n]}{FP_T[k,n]}\right| < \left|1-\dfrac{FP_T[k,n]}{FP_R[k,n]}\right| \\ \dfrac{FP_T[k,n]}{FP_R[k,n]} & \text{if } \left|1-\dfrac{FP_T[k,n]}{FP_R[k,n]}\right| < \left|1-\dfrac{FP_R[k,n]}{FP_T[k,n]}\right| \end{cases} \quad (30)$$

$FP_R[k,n]$ and $FP_T[k,n]$ represent the *Filtered Pitch Patterns* for the Reference and Test audio signals respectively. We calculate the mean and the variance of $D_p[k,n]$ over the critical bands domain as shown in (31) and (32).

$$\Omega_1[n] = \frac{1}{Z}\sum_{k=0}^{Z-1} D_p[k,n] \quad (31)$$

$$\Omega_2[n] = \frac{1}{Z}\sum_{k=0}^{Z-1}(D_p[k,n])^2 - (\Omega_1[n])^2 \quad (32)$$

We have observed that as the compression rate increases, both pitch patterns, $FP_R[k,n]$ and $FP_T[k,n]$, will have a greater resemblance to each other; their statistical measures will so. In this regards, the mean ($\Omega_1$) and the variance ($\Omega_2$) of $D_p[k,n]$ over the time domain will approximate to 1 and to 0 respectively, i.e. the closer $\Omega_1$ is to the unity and the closer $\Omega_2$ is to zero, the better quality the audio is.

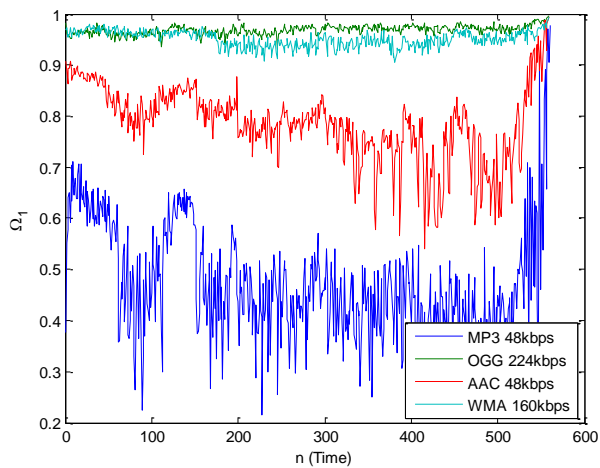

Fig. 5. Mean over the time domain

Fig. 5 illustrates $\Omega_1[n]$ for 4 audio samples and can be noticed that in average the values of $\Omega_1[n]$ tends to a higher value when the bitrate is increased. $\Omega_1[n]$ will not exceed the unity since $D_p[k,n]$ is bounded to the range [0:1].

On the other hand, Fig. 6 illustrates $\Omega_2[n]$ for 4 audio samples and can be concluded from it, that when the bitrate is high the variance of $D_p[k,n]$ is low, almost tending to zero. However low bitrate compressed audio signals have a high value of variance in comparison with those obtained from the high bitrate compressed audio signals.

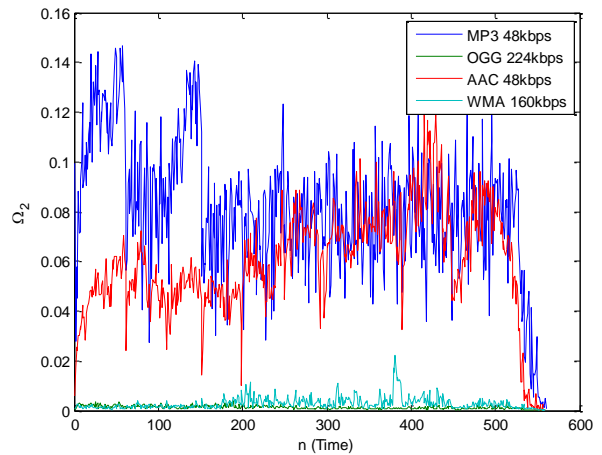

Fig. 6. Variance over the time domain

In base of (31) and (32), we compute the temporal average of $\Omega_1[n]$ and a scaled temporal average of $\Omega_2[n]$ as shown in (33) and (34).

$$I_1 = \frac{1}{N}\sum_{n=o}^{N-1}\Omega_1[n] \quad (33)$$

$$I_2 = \min\left\{\gamma \cdot \frac{1}{N}\sum_{n=0}^{N-1}\Omega_2[n],\ 1\right\} \quad (34)$$

$I_1$ and $I_2$ will be used as inputs in the fuzzification process. Consider that $\gamma$ was set to 10 in order to normalize $I_2$ to the range [0:1] since it was experimentally found that $I_2$ took values in the range [0-0.1].

*D. Fuzzy C-Means Clustering Approach*

Fuzzy C-Means (FCM) is an unsupervised method whose main target is to classify a data set into subsets (clusters) so that one piece of data could belong to two or more clusters. Elements within a cluster share certain degree of similarity with the other elements of the cluster. FCM is commonly used for statistical data analysis in many fields such as data mining, machine learning and pattern recognition [16]. This algorithm is based on the minimization of (35)

$$J_m = \sum_{i=1}^{L}\sum_{j=1}^{C} u_{ij}^{m} \|x_i - c_j\|^2 \quad (35)$$

Where $L$ represents the total number of elements to be clustered and $C$ represents the total number of clusters. Also assume $m = 2$.

In (35), $c_j$ represents the $d$-dimensional centroid of the cluster $j$; $x_i$ represents the $i^{th}$ element of the data set; $u_{ij}$ represents the degree of membership of $x_i$ in the cluster $j$ and $\|*\|$ represents the Euclidean distance between element $x_i$ and centroid $c_j$. Fuzzy clustering is carried out running an iterative process in which the function shown in (35) is to be minimized through the update of the membership degree matrix $u_{ij}$ and the centroids $c_j$.

Equation (36) shows the formula to compute the degree of membership.

$$u_{ij} = \frac{1}{\|x_i - c_j\|^{\frac{2}{m-1}} \cdot \sum_{k=1}^{C}\left(\frac{1}{\|x_i - c_k\|}\right)^{\frac{2}{m-1}}} \quad (36)$$

In (37) it is shown the process for computing the centroids of the clusters

$$c_j = \frac{\sum_{i=1}^{N} u_{ij}^{m} \cdot x_i}{\sum_{i=1}^{N} u_{ij}^{m}} \quad (37)$$

As mentioned above, the fuzzy partitioning is done iteratively until the stop condition defined in (38) occurs

$$\max_{ij}\left\{\left|u_{ij}^{(iteration)} - u_{ij}^{(iteration-1)}\right|\right\} < \varepsilon \quad (38)$$

where $\varepsilon$ represents a termination criterion [16].

In our experiments we set the number of clusters to six. From the data collected in the subjective tests, we randomly selected 84 audio samples to which measures $I_1$ and $I_2$ were computed and plotted as shown in Fig. 7.

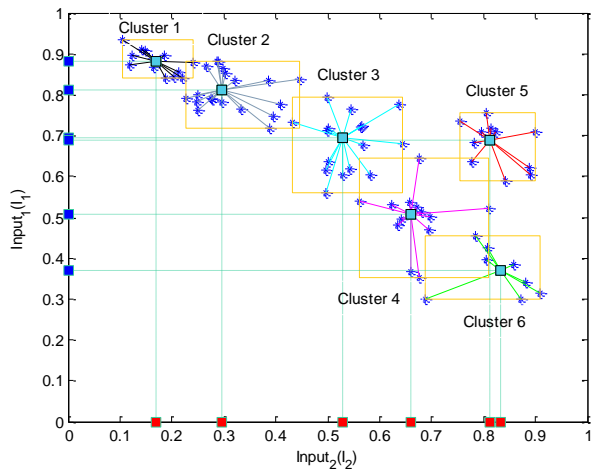

Fig. 7.    Clusters that resulted from FCM

We can infer from Fig.7 that some neighboring clusters are too close to each causing their projections have a high degree of overlapping. For example, clusters 3 and 5 (see Fig. 8) on the ordinate axis and clusters 5 and 6 (see Fig. 9) on the abscissa axis are characterized by a high crosspoint level, i.e. high degree of overlapping. Merging can be used in order to combine membership functions with high crosspoint level.

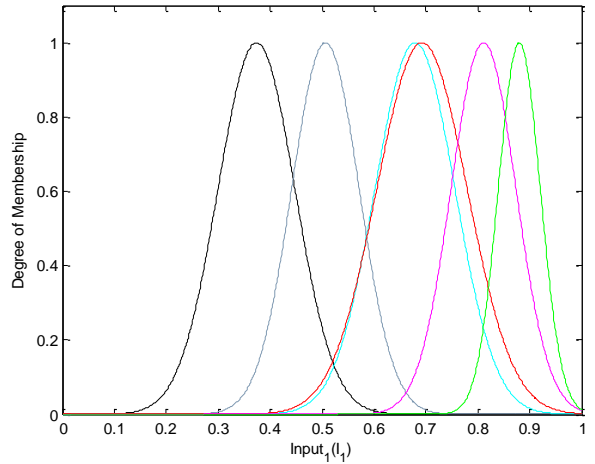

Fig. 8.    Membership Functions for Input 1

The idea is to make use of the centroids and the variances of the clusters, which reflect the actual data distribution, to generate appropriate membership functions for each input. Clusters are approximated as ellipses with its center being the clusters centroids and the lengths of axes decided by the corresponding variances of the clusters [17]. By projecting an ellipse on one axis will generate a symmetric membership function with its peak point being the cluster centroid as seen in Fig. 8 and Fig. 9.

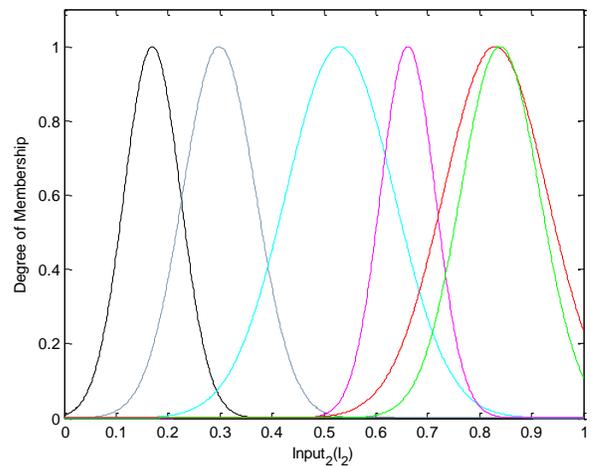

Fig. 9.    Membership Functions for Input 2

Variances for each cluster are calculated according to (39)

$$v_j = \lambda \cdot \frac{\sum_{i=1}^{N} u_{ij}^{m} \cdot (x_i - c_j)^2}{\sum_{i=1}^{N} u_{ij}^{m}} \qquad (39)$$

Assume that λ is a value that defines the degree of overlapping between neighboring clusters, i.e. represents the fuzziness of the data. In this paper λ was set to 1.5.

*E. Membership Functions Merging*

Merging of two membership functions occurs when they become sufficiently close to each other as exemplified above with Fig. 8 and Fig. 9. Two membership functions are fused when the supremum of their intersection exceeds a threshold [19]. If the means of the membership functions prior to fusion are $\mu_1$ and $\mu_2$, then the mean of the merged membership is calculated as shown in (40)

$$\mu = \frac{\mu_1 \cdot \sigma_1 + \mu_2 \cdot \sigma_2}{\sigma_1 + \sigma_2} \qquad (40)$$

Consider that $\sigma_1$ and $\sigma_2$ are the standard deviation values of the two membership functions.

$$\sigma^2 = \frac{\sigma_1^3 + \sigma_2^3}{\sigma_1 + \sigma_2} \qquad (41)$$

In (41) $\sigma^2$ represents the variance of the merged membership function.

Finally, ten membership functions resulted after the merging process, five for each input. We resume the clusters features in TABLE II.

TABLE II.  CLUSTERS FEATURES

| Cluster | Axis projection | Centroid | Variance |
|---|---|---|---|
| 1 | $I_1$ | 0.8799 | 0.0405 |
| 2 | $I_1$ | 0.8110 | 0.0625 |
| 3 | $I_1$ | 0.6862 | 0.0826 |
| 4 | $I_1$ | 0.5066 | 0.0651 |
| 5 | $I_1$ | 0.3738 | 0.0741 |
| 1 | $I_2$ | 0.1697 | 0.0543 |
| 2 | $I_2$ | 0.2986 | 0.0703 |
| 3 | $I_2$ | 0.5313 | 0.1040 |
| 4 | $I_2$ | 0.6618 | 0.0527 |
| 5 | $I_2$ | 0.8343 | 0.0903 |

*F. Fuzzification*

As a first step in fuzzy logic systems, real world input variables (crisp values) must be transformed into fuzzy values. Crisp values are fuzzified using input membership functions (MF).

The membership functions for both input variables are illustrated in Fig. 10 and Fig. 11. We used Gaussian membership functions because they facilitate obtaining smooth hypersurfaces which is closely related with soft decision.

The membership functions for the output are shown in Fig. 12 and four linguistic variables {P, M, G, H} were conceived, where P accounts for Poor Quality, M for Medium Quality, G for Good Quality and H for High Quality. Gaussian MFs were also used. In this case, the output MFs represent the categories in which the audio quality can be classified.

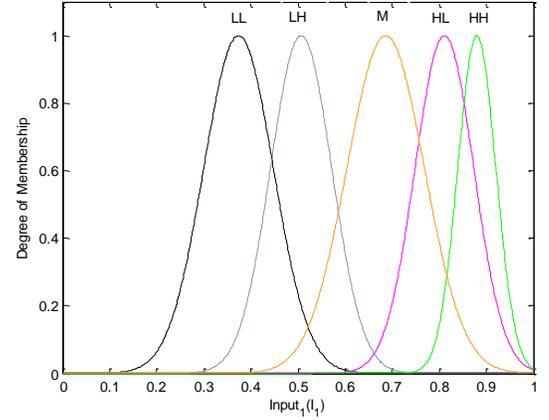

Fig. 10.    Membership Functions for Input 1

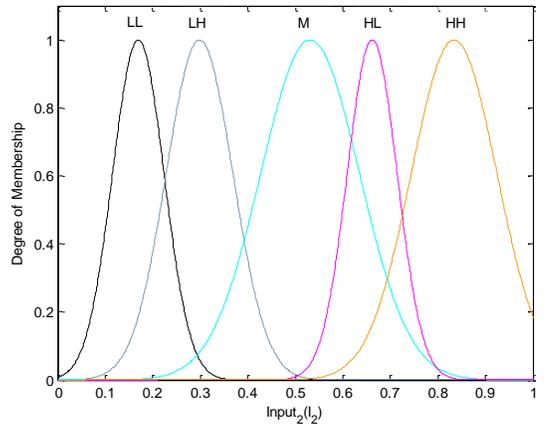

Fig. 11.    Membership Functions for Input 2

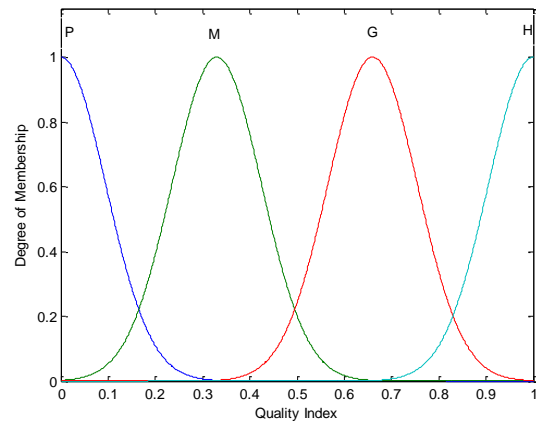

Fig. 12.    Output Membership Function

In order to combine the fuzzy relations we used the MAX-MIN composition method that used by Zadeh [20] in his original paper about approximate reasoning using if-then rules as features of natural language.

### G. Fuzzy Rules

The rule base is composed of 25 fuzzy decision rules and has been defined in TABLE III according to our experiments and observations.

TABLE III. FUZZY RULE

| Input 1 \ Input 2 | LL | LH | M | HL | HH |
|---|---|---|---|---|---|
| LL | M(1) | P(2) | P(3) | P(4) | P(5) |
| LH | M(6) | M(7) | P(8) | P(9) | P(10) |
| M | G(11) | G(12) | M(13) | P(14) | P(15) |
| HL | H(16) | H(17) | G(18) | M(19) | P(20) |
| HH | H(21) | H(22) | H(23) | M(24) | P(25) |

In rule 22, e.g. when $I_1$ is HH (High-High) and $I_2$ is LH (Low-High) then the output is H (High). In another case, in rule 27 if $I_1$ is M (Medium) and $I_2$ is HL (High-Low), the output is classified as P (Poor). The fuzzy rules represent all the possible combinations for Input 1 and Input 2 and the corresponding output.

### H. Defuzzification

After performing the fuzzy composition through the use of membership functions and the Fuzzy Rules, we obtain a fuzzy set as shown in Fig. 13

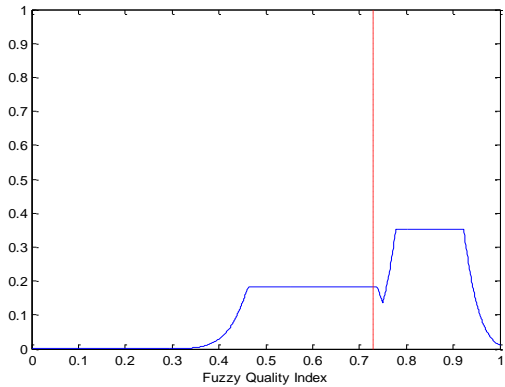

Fig. 13. Defuzzification Process

We used the centroid approach to convert the fuzzy values into crisp values. The centroid position is highlighted with red color in Fig. 13. The resulting value of this stage is called Fuzzy Quality Index, which represents the perceived audio quality of audio signals.

## V. ANN TRAINING

We used a Three-Layer Perceptron Neural Network; with 12 linear neurons at the input, 12 sigmoidal neurons at the hidden layer and one single neuron for the output. We used Resilient Propagation [12] for training the network.

## VI. RESULTS

In order to test the performance of our metric, we selected four audio codecs; LAME MP3, OGG Vorbis, MPEG4-AAC and WMA. We processed 10 audio files at different bitrates using the aforementioned codecs. We computed the mean of the FQI values that were obtained for the ten audio files and plotted the results in Fig. 14.

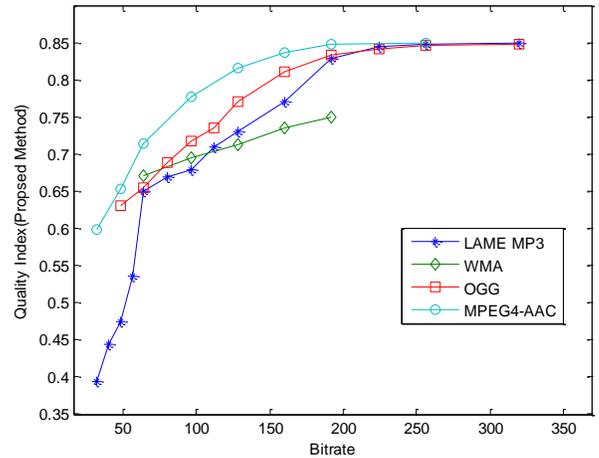

Fig. 14. Results of the proposed metric (FQI)

It can be said based on the results of our metric, that AAC has the best performance for the entire range of bitrates. OGG Vorbis performs slightly better than LAME MP3. WMA performs better than OGG and MP3 for low bitrates; however it does not for higher bitrates.

TABLE IV. CORRELATION BETWEEN FQI AND SDG

| Codec | LAME MP3 | OGG | MPEG4-AAC | WMA |
|---|---|---|---|---|
| Pearson Corr. Coefficient | 0.8693 | 0.9604 | 0.9704 | 0.9221 |

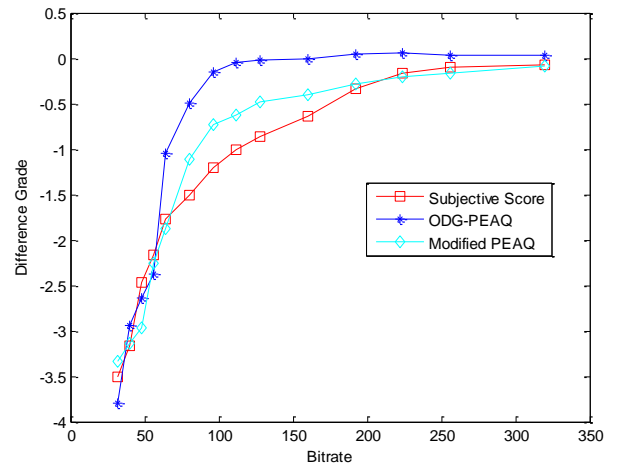

Fig. 15. ODG for a MP3 Files Dataset

TABLE IV shows the correlation coefficients between our metric and the Subjective Difference Grade. Note that by itself the FQI metric can achieve a high degree of correlation with respect to the subjective opinion.

With the same audio dataset that was used to compute the FQI in Fig.14, we computed the ODG for the original PEAQ version as well as for the Modified PEAQ which incorporates our metric. Fig. 15 shows the performance of the Modified PEAQ version with respect to the original PEAQ version using only MP3 coded files. On the other hand Fig.16 shows the results for both PEAQ versions using only OGG audio files.

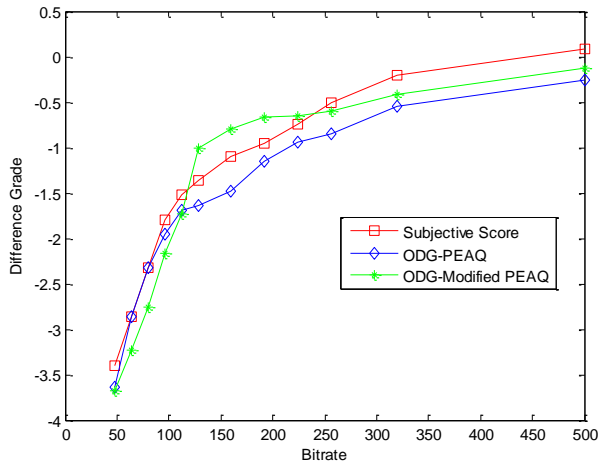

Fig. 16. ODG for a OGG Files Dataset

For both experiments, correlation coefficients as well as the Mean Square Error are shown in TABLE IV.

TABLE V. COMPARISON BETWEEN PEAQ AND MODIFIED PEAQ

| Codec | Modified PEAQ | | Original PEAQ | |
|---|---|---|---|---|
| | Pearson Corr.Coef | MSE | Pearson Corr.Coef | MSE |
| LAME MP3 | 0.9762 | 0.0732 | 0.947 | 0.3643 |
| OGG | 0.9760 | 0.0835 | 0.9951 | 0.0636 |

In the first experiment (MP3 files), our method had achieved better correlation than the original PEAQ algorithm, in addition the MSE is lower. In the second case the original PEAQ algorithm had achieved better correlation than ours but MSE values are almost the same.

## VII. CONCLUSIONS

We have observed that by incorporating the proposed metric to the original PEAQ model; in general we have achieved better correlation between the ODG and the Subjective Difference Grade. Even in the cases where there was no better degree of correlation, the total error is almost the same. As we mentioned before, the Fuzzy Quality Index has by itself good consistency with respect to the values of bitrates but it is worth saying that the proposed metric is not infallible since there are still a wide range of audio codecs that have not been tested in the experiments. The fact of using twelve hidden neuron rather than three (used in the original PEAQ version) could have improved the overall performance by further minimizing the error.